\documentclass[11pt]{article}
\usepackage{amsmath,amsxtra,amssymb,latexsym, amscd}
\usepackage[mathscr]{eucal}

\makeindex
\begin{document}
\parindent=1.05cm 
\setlength{\baselineskip}{12truept} \setcounter{page}{1}
\makeatletter
\renewcommand{\@evenhead}{\@oddhead}
\renewcommand{\@oddfoot}{}
\renewcommand{\@evenfoot}{\@oddfoot}
\renewcommand{\thesection}{\arabic{section}.}
\renewcommand{\thesubsection}{\thesection\arabic{subsection}.}
\renewcommand{\theequation}{\thesection\arabic{equation}}
\@addtoreset{equation}{section}
\begin{center}
\vspace{10cm}
{\bf THE CONTRIBUTION OF EFFECTIVE QUANTUM GRAVITY TO HIGH ENERGY SCATTERING IN THE FRAMEWORK OF MODIFIED PERTURBATION THEORY AND ONE LOOP APPROXIMATION}\\
\vspace{0.5cm}
\small{
Nguyen Suan Han$^{1,2}$\footnote{Email:suanhan@hus.edu.vn, suan@theor.jinr.ru, lienbat76@gmail.com}, Do Thu Ha$^{2}$,Nguyen Nhu Xuan$^{3}$\\
\vspace{0.5cm}
{$^1$\it Bogoliubov Laboratory of Theoretical Physics, JINR, Dubna, 141980, Russia.} \\
{$^2$\it Department of Theoretical Physics, Hanoi University of Science, Hanoi, Vietnam.} \\
{$^3$\it Department of Physics, Le Qui Don University, Hanoi, Vietnam.}\\}
\end{center}
\vspace{0.5cm}
\baselineskip=18pt
\bigskip
\textbf{Abstract}: The asymptotic behavior of the scattering amplitude for two scalar particles at high energies with fixed momentum transfers is studied. The study is done within the effective theory of quantum gravity based on quasi-potential equation. By using the modified perturbation theory, a systematic method is developed to find the leading eikonal scattering amplitudes together with corrections to them in the one-loop gravitational approximation. The relation is established and discussed between the solutions obtained by means of the operator and functional approaches applied to quasi-potential equation. The first non-leading corrections to the leading eikonal amplitude are found.
\vspace{0.5cm}
\newpage
\section{Introduction}
\indent The asymptotical behavior of the scattering amplitude at high energy for all types of interactions including gravitational interaction is one of the central problems of elementary particle physics. From the standard method of quantum field theory it follows that calculations based on perturbation theory are suitable when the energy of individual particles is not very high and the effective coupling constant is not very large. When the energy is increased the effective coupling constant also increases so that the corrections calculated by perturbation theory play a crucial role. The gravitational scattering occurs at energies $\sqrt{s}=2E<< M_{PL}-$ and is described by “effective field theory”, where $s$ is the squared energy of the center of mass, $M_{PL}$ is the Planck mass, $G$ is the universal gravitational constant, which is characterized by the effective coupling constant $\alpha_G=Gs/\hbar\geq 1$ and makes any simple perturbation expansion unwarranted. Comparison of the results of different approaches tackling this problem has shown to coincide in the leading order of approximation that has a semi-classical effective metric interpretation, while the most of them fail in providing the non-leading terms [1-10]. The determination of these corrections to gravitational scattering is currently an open problem. These corrections play a crucial role in such problems as strong gravitational forces near the black hole, string modification of the gravitational theory and some other effects of quantum gravity [2,3,6,7].

\indent In the framework of standard field theory and the high-energy scattering, the different methods have been developed to investigate the asymptotic behavior of individual Feynman diagrams and their subsequent summation. The calculations of the eikonal diagrams in the case of gravity run similarly to the analogous calculations in QED. The calculations of eikonal capture the leading behavior of each order in perturbation theory, but the sum of leading terms is subdominant to the neglected term by this approximation. The reliability of eikonal amplitude for gravity is uncertain. An approach that has probed the first of these features with some success is the one that based on the reggeized string exchange amplitudes with subsequent reduction to the gravitational eikonal limit including the leading order corrections [11]. In articles [12-14] the high energy scattering amplitude of two “nucleons” in the quantum gravity is constructed by extending the functional integration method [15-20] which has been used effectively in quantum electrodynamics [21-27]. A straight-line path approximation was used to calculate the functional integrals which occur. It is shown that in the limit of asymptotically high energy $s\rightarrow\infty$ and at fixed $t$ - momentum transfer, the elastic scattering amplitude of two “nucleons” has the form of a Glauber representation  with an eikonal function depending on the energy. A similar result is obtained by the “shock-wave method” proposed by ’tHooft[1], by the method of eﬀective topological theory in the Planck limit proposed by Verlinde and Verlinde [8] and by summation of Feynman diagrams in the eikonal approximation [9]. The main advantage of the proposed approach over the other approaches known is the possibility to perform calculations in a compact form. In doing so, the correct structure of the Green’s function and amplitudes etc. is not destroyed by approximations in the process of the calculations.

\indent It should be emphasized that in the framework of functional integrals the modified perturbation theory proposed by Fradkin in [16,17] leads to results that go well beyond those of ordinary perturbation theory. This method is especially fruitful in quantum electrodynamics since even in the first approximation of the modified perturbation theory the contribution of both the virtual and the real soft photons can be completely summed. In particular, it was shown that the first approximation of the modified perturbation theory is already sufficient to obtain the correct asymptotic behavior for the Green's functions and the cross sections of processes in quantum electrodynamics [16-18] in the so-called log-log approximation. These general results allow to solve the problem of finding the quantum Green's functions and cross sections of processes with allowance for the radiative corrections in electrodynamics in the presence of a real external field without resorting to perturbation theory.

\indent The purpose of the present paper is to develop a systematic scheme based on modified perturbation theory to find the correction terms to the leading eikonal amplitude by solving the Logunov –Tavkhelidze quasi-potential equation [29-40]. In spite of the lack of a clear relativistic covariance, the quasi-potential method keeps all information about properties of scattering amplitude which could be received from the general principle of quantum field theory [29]. Therefore, at high energies one can investigate analytical properties of the scattering, its asymptotic behavior and some regularities of a potential scattering etc., exactly, as it has been done in the usual S-matrix theory. The choice of this approach is dictated also by the following reasons: i/ in the framework of the quasi-potential approach, the eikonal amplitude has a rigorous justification in quantum field theory [33]; ii/ in the case of smooth potentials, it was shown that a relativistic quasi-potential and the Schrodinger equations lead to qualitatively identical results [32,36].

\indent This paper is organized as follows: In the second section we briefly introduce the Logunov-Tavkhelidze quasi-potential equation, and rewrite this equation on a mass shell in the operator form, after defining an appropriate pseudo-differential operator. The solution of this equation is presented in an exponent form which is favorable to modify the perturbation method in the framework of functional integrals in the third section. The asymptotic behavior of scattering amplitude at high energies and fixed momentum transfer is also considered and a systematic scheme of finding the leading eikonal scattering amplitudes and its corrections are constructed in the fourth section. The lowest – order approximation of the modified perturbation theory is the leading eikonal scattering amplitude in the linearized quantum gravity. The first correction to leading eikonal amplitude is also found. The quasi-potential used here corresponds to virtual particles between two “nucleons”. The fifth section is devoted to calculating the one-loop leading eikonal scattering amplitude and its first correction to leading amplitude in the effective theory of quantum gravity. The new results here for the non-analytic and non-local contributions were discussed and included in the calculation of the behavior scattering amplitude by using the Newtonian potential with low-energy. The leading one loop gravitational corrections was obtained from S-matrix in quantum gravity. Finally, we discuss the obtained results and possible generalization of this approach.

\section{Two-particle quasi-potential equation in an operator form}

\indent For simplicity, we shall first consider the elastic scattering of two scalar “nucleons” with the interaction Lagrangian $L_{int}=g\varphi^{2}(x)\phi(x)$. The results will be generalized to the case of a scalar nucleon interacting with a neutral vector and a graviton fields later. For two scalar particle scattering amplitude the quasi-potential equation with local quasi-potential has the form [37]:
$$T({\vec{p}},{\vec{p}'};s)=gV({\vec{p}}-{\vec{p}'};s)
+g\int  d{\vec{q}} V({\vec{p}}-{\vec{q}};s) K({\vec{q}}^{2},s)
T({\vec{q}},{\vec{p'}};s), \eqno(2.1)$$
where
$K({\vec{q}}^{2},s)=\frac{1}{\sqrt{q^2+m^2}}\frac{1}{q^2+m^2-\frac{s}{4}-i\varepsilon}
$ is the kernel, $s=4E^2=4({\vec{p}}^{2}+m^{2})=4({\vec{p'}}+m^{2})$ is the energy and ${\vec{p}}, {\vec{p'}}$ are the relativistic momentums of two particles in the center of the mass reference frame in the initial and final states respectively. Equation (2.1) is one of the possible generalizations of the Lippman-Schwinger equation for the case of relativistic quantum field theory. The quasi-potential  in Eq. (2.1) is a complex function of the energy and relativistic momenta. The quasi-potential equation simplifies considerably if $V(r,s)$ is a function that depends only of the relative momenta and the total energy i.e. if the quasi-potential is local\footnote{Since the total energy as an external parameter of this equation, the “local” here has direct meaning and it can appear in a three-dimensional $\delta$ function in the quasi-potential in the coordinate representation}. The existence of the local quasi-potential has been well proven rigorously in the weak coupling case [35] and a method has been specified for constructing it. The local potential constructed in this manner gives a solution of Eq.(2.1) that is equal to the physical amplitude on the mass shell [29,30,32,35].\\
Performing the Fourier transformations
$$
V({\vec{p}}-{\vec{p'}};s)=\frac{1}{(2\pi)^{3}}\int d{\vec{r}}
e^{i({\vec{p}}-{\vec{p}'}){\vec{r}}} V({\vec{r}};s),\eqno (2.2)$$
$$
T({\vec{p}},{\vec{p'}};s)=\int d{\vec{r}} d {\vec{r'}}
e^{i({\vec{p}}{\vec{r}}-{\vec{p'}}{\vec{r'}})}
T({\vec{r}},{\vec{r'}};s),$$
and substituting Eq.(2.2) into Eq.(2.1), we obtain
$$
T(\vec{r}, \vec{r}' ; s)=\frac{g}{(2\pi)^{3}}V(\vec{r}; s)\delta^{(3)}(\vec{r}- \vec{r}')+ $$
$$\frac{g}{(2\pi)^{3}}\int\int d\vec{q} K(q^{2}; s)
V(\vec{r}; s) e^{-\vec{q}\vec{r}} \int d\vec{r}'' e^{i\vec{q}\vec{r}''} T(\vec{r}'',\vec{r}' ; s) \eqno (2.3)
$$
Introducing the representation
$$
T({\vec{r}},{\vec{r}'};s)= \frac{g}{(2\pi)^{3}}
V({\vec{r}};s)F({\vec{r}},{\vec{r'}};s), \eqno(2.4)$$
we find
$$
F({\vec{r}},{\vec{r'}};s)=\delta^{(3)}
({\vec{r}}-{\vec{r'}})+\frac{g}{(2\pi)^{3}}\int d{\vec{q}}
K({\vec{q}}^{2};s) e^{-i{\vec{q}}{\vec{r}}}\times
$$
$$ \times \int
d {\vec{r}''}e^{i{\vec{q}}{\vec{r}''}}
V({\vec{r}''};s)F({\vec{r}''},{\vec{r'}};s). \eqno(2.5)
$$
Defining the pseudo-differential operator $\widehat{L_{r}}=K( - {\bf{{\nabla_{r}}}}^{2};s)$, then
$$
K({\vec{r}};s)=\int d{\vec{q}}
K({\vec{q}}^{2};s)e^{-i{\vec{q}}{\vec{r}}}=K( -\nabla_{r}; s) \int
d{\vec{q}} e^{-i{\vec{q}}{\vec{r}}} = \widehat{L_{r}}
(2\pi)^{3}\delta^{(3)}({\vec{r}}). \eqno(2.6)
$$
After some simple transformations, Eq.(2.3) is rewritten in an operator form as:
$$
F({\vec{r}},{\vec{r'}};s)=\delta^{(3)}({\vec{r}}-{\vec{r'}})
+g\widehat{L_{r}}\bigl[V({\vec{r}},s)F({\vec{r}},{\vec{r'}},s)\bigl].
\eqno(2.7)
$$
Eq. $(2.7)$ is the operator form of the Logunov – Tavkhelizde equation[37].\\
\indent Within the framework of the quasi-potential approach, the potential is defined by expanding it into infinite series in order of the interaction constant $g$. It corresponds to the expansion of perturbation amplitudes on the mass shell. The approximate solution of Eq.(2.5) is found in the lowest order of the quasi-potential. Using this method, the relativistic eikonal expression of the scattering amplitude was found in quantum field theory with large energy and small momentum transfer [33].
\section{Modified perturbation theory}
\indent In quantum field theory, scattering problems are mostly solved by perturbation expansion. However, in some cases, in order to solve the problem smoothly, one can improve this method in an expansion that is called the modified perturbation method proposed by Fradkin\footnote{The
interpretation of the perturbation theory from the view-point of the
diagrammatic technique is as follows. The typical Feynman
denominator of the standard perturbation theory is of the form
$(A)$: $(p+\sum q )^{2}+m^{2}-i\varepsilon= p^{2}+m^{2}+2p\sum q
+(\sum q)^{2}$, where $ p $ is the external momentum of the scalar (
spinor ) particle, and the $ q's $ are virtual momenta of radiation
quanta. The lowest order approximation $ (A) $ of modified theory is
equivalent to summing all Feynman diagrams with the replacement: $
(\sum q)^{2}=\sum (q)^{2} $ in each denominator $(A)$. The modified
perturbation theory thus corresponds to a small correlation of the
radiation quanta: $ \bf{q_{i}}\bf{q_{j}}=0$ and is often called the
$\bf{q_{i}}\bf{q_{j}}$-approximation. In the framework of functional
integration this approximation is called the straight-line path
approximation i.e high-energy particles move along Feynman paths,
which are practically rectilinear.} in the framework of functional integrals [16-17], which goes well beyond the ordinary perturbation theory. Apply it to Eq.(2.7), we can write the solution of this equation in the symbolic form
$$
F({\vec{r}},{\vec{r'}};s)=\frac{1}{(2\pi)^3}\int d {\vec{k}}
\exp{\biggl[W({\vec{r}};{\vec{k}};s)\biggl]}
e^{-i{\vec{k}}({\vec{r}}-{\vec{r'}})}. \eqno(3.1)
$$
Substituting Eq.(3.1) into Eq.(2.7), we obtain an equation for the function $W(\vec{r},\vec{k};s)$
$$
\exp{W({\vec{r}};{\vec{k}};s)}=1+g \widehat{L_{r}}\left\{V({\vec{r}},s)
\exp \left[W(\vec{r},\vec{k};s)-i\vec{k}\vec{r}\right]\right\}
e^{i{\vec{k}}{\vec{r}}}. \eqno (3.2)
$$
Using the idea of the modified perturbation theory in exponent function, we can write the function $W(\vec{r},\vec{k};s)$ as an expansion in series in the coupling constant $g$
$$
W({\vec{r}};{\vec{k}};s)= \sum_{n=1}^{\infty} g^{n}
W_{n}({\vec{r}};{\vec{k}};s). \eqno(3.3)
$$
Substituting Eq.(3.3) into Eq.(3.2) and using Taylor expansion, the
l.h.s of Eq.(3.2) is rewritten as
$$1+\sum_{n=1}^\infty g^nW_n+\frac{1}{2!}\left(\sum_{n=1}^\infty
g^nW_n\right)^2+\frac{1}{3!}\left(\sum_{n=1}^\infty
g^nW_n\right)^3+\ldots,\eqno(3.4)$$
and the r.h.s of Eq.(3.2) has the form
$$
1+g\Biggl\{\hat{L}_r\left[V(\mathbf{r};s)\left(1+\sum_{n=1}^\infty
g^nW_n +\frac{1}{2!}\left(\sum_{n=1}^\infty
g^nW_n\right)^2+\frac{1}{3!}\left(\sum_{n=1}^\infty
g^nW_n\right)^3+\ldots\right)\right]+$$
$$
+V(\mathbf{r};s)\left[1+\sum_{n=1}^\infty g^nW_n
+\frac{1}{2!}\left(\sum_{n=1}^\infty
g^nW_n\right)^2+\frac{1}{3!}\left(\sum_{n=1}^\infty
g^nW_n\right)^3+\ldots\right]K(\mathbf{k};s)\Biggl\}.\eqno(3.5)
$$
Using Eq.(3.4) and Eq.(3.5) and comparing with the both sides of Eq.(3.2) in order of the coupling constant $g$, we derive the first, second, and third order approximation terms, respectively
$$
W_{1}({\vec{r}};{\vec{k}};s)= \int d{\vec{q}}V({\vec{q}};s)
 K[({\vec{k}}+{\vec{q}})^{2};s] e^{-i{\vec{q}}{\vec{r}}}; \eqno(3.6)
$$
$$
W_{2}({\vec{r}};{\vec{k}};s)= -\frac{W_{1}^{2}({\vec{r}};{\vec{k}};s)}
{2!}+$$
$$
\frac{1}{2} \int d{\vec{q}}_{1} d{\vec{q}}_{2}
V({\vec{q}}_{1};s)V({\vec{q}}_{2};s) K[({\vec{k}}+ {\vec{q}}_{1}+
{\vec{q}}_{2})^2;s]\times
$$
$$\times [ K ({\vec{k}}+ {\vec{q}}_{1}; s)
+K[{\vec{k}}+{\vec{q}}_{2};s ]] e^{
-i{\vec{q}}_{1}{\vec{r}}-i{\vec{q}}_{2}\vec{r}}; \eqno(3.7)
$$
$$
W_{3}({\vec{r}};{\vec{k}};s)=
-\frac{W_{1}^{2}({\vec{r}};{\vec{k}};s)}{3!}+ \int
d{\vec{q}}_{1}d{\vec{q}}_{2}d{\vec{q}}_{3}
V({\vec{q}}_{1};s)V({\vec{q}}_{2};s)V({\vec{q}}_{3};s)
K[({\vec{k}}+{\vec{q}}_{1})^{2};s] \times$$
$$\times K[({\vec{k}}+
{\vec{q}}_{1}+{\vec{q}}_{2})^{2};s] K[({\vec{k}}+
{\vec{q}}_{1}+{\vec{q}}_{2}+{\vec{q}}_{3})^{2};s ]
e^{-i{\vec{q}}_{1}{\vec{r}}-i{\vec{q}}_{2}
{\vec{r}}-i{\vec{q}}_{3}{\vec{r}}}. \eqno(3.8)
$$
etc.\\
Restricting the treatment to only $W_1(\vec{r},\vec{k};s)$ instead of $W(\vec{r},\vec{k};s)$ in Eq.(3.1)
we obtain from Eqs.(3.1), (2.4) and (2.2) approximate expression for the scattering amplitude
$$
T_{1}({\vec{p}},{\vec{p'}};s)=\frac{g}{(2\pi)^{3}} \int d{\vec{r}}
e^{i({\vec{p}}-{\vec{p'}}){\vec{r}}} V({\vec{r}},s)
e^{gW_{1}({\vec{r}},{\vec{p}},s)}. \eqno(3.9)
$$
To establish the meaning of this approximation, we expand $T_1(\vec{r},\vec{k};s)$ in series in the coupling constant $g$:
$$
T_{1}^{(n+1)}({\vec{p}},{\vec{p'}};s)=\frac{g^{n+1}}{n!}\int
d{\vec{q}}_{1}...d{\vec{q}}_{n} V({\vec{q}}_{1};s)....V({\vec{q}}_{n};s)
$$
$$\times V({\vec{p}}-{\vec{p'}} -\sum_{i=1}^{n}
q_{i};s) \prod_{i=0}^{n} K[({\vec{q}}_{i}+{\vec{p'}})^{2};s].
\eqno(3.10)
$$
and compare it with the $ (n+1)-th $ iteration term of
exact Eq. $(2.1)$
$$
T^{(n+1)}({\vec{p}},{\vec{p'}};s)=\int
d{\vec{q}}_{1}...d{\vec{q}}_{n}V({\vec{q}}_{1};s)... V({\vec{q}}_{n};s)
V({\vec{p}}-{\vec{p'}} -\sum_{i=1}^{n} q_{i};s)\times
$$
$$ \sum_{p}K[({\vec{q}}_{1}+{\vec{p'}})^{2};s]
K[({\vec{q}}_{1}+ {\vec{q}}_{2}+{\vec{p'}})^{2};s]... K[(
\sum_{i=1}{\vec{q}}_{i}+{\vec{p'}})^{2};s], \eqno(3.11)
$$
where $\sum_{p}$ is the sum over the permutations of the momenta
${\vec{p}}_{1}$ ,${\vec{p}}_{2}...$ ${\vec{p}}_{n}$. \\
It is readily seen from $(3.10)$ and $(3.11)$ that our approximation in the case of the
Lippmann-Schwinger equation is identical with the ${{\vec{q}}_{i}}{{\vec{q}}_{j}}=0$ approximation, in accordance with which terms of the type ${{\vec{q}}_{i}}{{\vec{q}}_{j}}=0, i\neq j$ are ignored in the "nucleon  propagators".
\section{Asymptotic behavior of the scattering amplitude at high energy}
\indent In this section the solution of the Logunov-Tavkhelidze quasi-potential equation obtained in the previous section for the scattering amplitude can be used to find the asymptotic behavior amplitude in the high energy $ s\rightarrow \infty$ and ﬁxed $t$ - momentum transfer. In asymptotic expansions, we shall retain both the principal term and the next order term, using the formula
$$
e^{W({\vec{r}},{\vec{p'}}; s)}= e^{W_{1}({\vec{r}},{\vec{p'}};
s)}\biggl [1+g^{2}W_{2}({\vec{r}},{\vec{p'}};s)+...\biggl],
\eqno(4.1)
$$
where $W_{1}(\vec{r},\vec{k};s)$ and $W_{2}(\vec{r},\vec{k};s)$ were determined by Eqs.(3.6) and (3.7), respectively.\\
Take the $ z $ axis along the $({\vec{p}}+{\vec{p'}})$ momentum of the incident particles and use Mandelstam variables, we have
$$
\vec{\Delta}_\perp.\vec{n}_z=0, t=-\vec{\Delta}_{\perp}^{2}, \vec{p}-\vec{p'}=\vec{\Delta}_{\perp}. \eqno(4.2)
$$
Noting
$$
K({\vec{p}}+{\vec{p'}};s) =\frac{1}
{\sqrt{({\vec{p}}+{\vec{p'}})^{2}+m^{2}}}\frac{1}
{({\vec{p}}+{\vec{p'}})^{2}-\frac{s}{4}+m^{2}-i\varepsilon}\Biggr|_{s\rightarrow\infty,t-fixed}$$
$$
=\frac{2}{s(q_{z}^{2}-i\varepsilon)} [1-\frac{ 3q_{z}^{2}+
{\vec{q}_{\perp}}^{2} + {\vec{q}_{\perp}}{\bf{\triangle}_{\perp}}}
{\sqrt{s}(q_{z}-i\epsilon)}] +
 O(\frac{1}{s^{2}}), \eqno(4.3)
$$
Substitute Eq.(3.9) into Eq.(3.5) and Eq.(3.6), we get
$$
W_{1}=\biggl (\frac{W_{10}}{s}\biggl ) + \biggl
(\frac{W_{11}}{s\sqrt{s}}\biggl ) + O \biggl(\frac{1}{s^{2}}\biggl
); \eqno(4.4)
$$
$$
W_{2}=\biggl (\frac{W_{20}}{s^{2}\sqrt{s}}\biggl ) +O \biggl
(\frac{1}{s^{3}}\biggl ), \eqno(4.5)
$$
where the $W_{10},W_{11}$ and $W_{20}$ terms are calculated in detail in Ref. [14]
$$
W_{10}=2\int d{\vec{q}} V({\vec{q}};s) \frac{e^{i{\vec{q}}{\vec{r}}}}
{(q_{z}^{2}-i\varepsilon)^{2}} = 2i\int_{-\infty}^{z} dz'
V(\sqrt{{\vec{q}_{\perp}}^{2}+z'^{2}};s); \eqno (4.6) $$

$$
W_{11}=-2 \int d{\vec{q}}V({\vec{q}};s) e^{-i{\vec{q}}{\vec{r}}} \frac{
3q_{z}^{2}+ {\vec{q}_{\perp}}^{2} +
{\vec{q}_{\perp}}{\bf{\triangle}_{\perp}}} {(q_{z}-i\epsilon)^2} =$$
$$ =-6 V(\sqrt{{\vec{q}_{\perp}}^{2}+z'^{2}};s) + 2(
-{\bf{{\nabla}_{\perp}}}^{2}-i\vec{q}_{\perp}{\bf{{\nabla}_{\perp}}})
\int_{-\infty}^{z} dz' V(\sqrt{{\vec{q}_{\perp}}^{2}+z'^{2}};s);
\eqno(4.7)$$

$$
W_{20}=-4 \int d{\vec{q}}_{1}d{\vec{q}}_{2}
e^{-i({\vec{q}}_{1}+{\vec{q}}_{2}){\vec{r}}}
V({\vec{q}}_{1};s)V({\vec{q}}_{2};s)
\frac{3q_{1z}q_{2z}+{\vec{q}}_{1\perp}{\vec{q}}_{2\perp}}
{(q_{1z}-i\varepsilon)(q_{2z}-i\varepsilon)(q_{1z}+q_{2z}-i\varepsilon)}$$
$$= -4i\left\{ 3\int_{-\infty}^{z} dz'
V^{2}(\sqrt{{\vec{q}_{\perp}}^{2}+z'^{2}};s)+ \left[
{\bf{{\nabla}_{\perp}}}\int_{-\infty}^{z'} dz''
V^2(\sqrt{{\vec{q}_{\perp}}^{2}+z''^{2}};s)\right]^{2}\right\}.
\eqno(4.8)
$$
In the limit $s\rightarrow\infty$ and $(t/s)\rightarrow 0$, $W_{10}$ makes the main contribution, and the remaining terms are corrections. Therefore, the function $W$ can be represented by means of the expansion (4.1) where $W_{10}, W_{11}$ and $W_{20}$ are determined by Eqs.(4.6)-(4.8),
respectively. The asymptotic behavior scattering amplitude can be written in the following form
$$
{\left.T(s;t)\right|_{\begin{array}{*{10}{c}}
{s\to \infty}\\
{t - fixed}
\end{array}}}=\frac{g}{(2\pi)^{3}}\int d^{2}\vec{r}_{\perp}dz
e^{i{\bf{\Delta_{\perp}}}{\bf{r_{\perp}}}}V(\sqrt{\vec{r}^{2}+z^{2}};s)\times$$
$$\times \exp\biggl(g\frac{W_{10}}{s}\biggl)
\biggl(1+g\frac{W_{11}}{s\sqrt{s}}+g^{2}\frac{W_{20}}{s^{2}\sqrt{s}}+...\biggl).
\eqno(4.9)
$$

Substituting Eqs.$(4.6)-(4.8)$ into Eq.$(4.9)$ and making calculations, at
the $s\rightarrow\infty$ and t- fixed momentum transfer, we finally obtain

$$
{\left.{T_{Scalar}^{\left( 0 \right)}(s;t)} \right|_{\begin{array}{*{10}{c}}
{s\to \infty }\\
{t - fixed}
\end{array}}}=-\frac{is}{2(2\pi)^{3}} \int
d^{2}\vec{r}_{\perp}e^{i{\vec{\Delta}_{\perp}}{\vec{r}_{\perp}}}
\Biggl\{e^{\bigl[\frac{2ig}{s}\int_{-\infty}^{\infty}dz
V(\sqrt{\vec{r}^{2}+z^{2}};s)\bigl]}-1 \Biggl\}-
$$

$$
-\frac{6g^{2}}{(2\pi)^{3}s\sqrt{s}} \int d^{2}\vec{r}_{\perp}
e^{i{\vec{\Delta}_{\perp}}{\vec{r}_{\perp}}} \exp\biggl[\frac{2ig}{s}
\int_{-\infty}^{\infty} dz
V(\sqrt{\vec{r}_{\perp}^{2}+z^{2}};s)\biggl] \int_{-\infty}^{\infty}
dz V^2(\sqrt{\vec{r}_{\perp}^{2}+z^{2}};s)-$$
$$
-\frac{ig}{(2\pi)^{3}\sqrt{s}}\int d^{2}\vec{r}_{\perp}
e^{i{\vec{\Delta}_{\perp}}{\vec{r}_{\perp}}} \times$$
$$\times\int_{-\infty}^{\infty}dz \Biggl \{\exp \biggl[\frac{2ig}{s}
\int_{z}^{\infty} dz'V(\sqrt{\vec{r}_{\perp}^{2}+z'^{2}};s) \biggl] -
\exp{\biggl[\frac{2ig}{s} \int_{-\infty}^{\infty} dz'
V(\sqrt{\vec{r}_{\perp}^{2}+z'^{2}};s)\biggl]}\Biggl\}\times$$
$$\times \Biggl\{\int_{z}^{\infty} dz'
{\bf{\nabla_{\perp}}}^{2}V(\sqrt{\vec{r}_{\perp}^{2}+z'^{2}};s)-
\frac{2ig}{s}
\biggl[\int_{z}^{\infty}dz'{\vec{\nabla}_{\perp}}V(\sqrt{\vec{r}_{\perp}^{2}+z^{2}};s)
\biggl]^{2}\Biggl\}$$
$$-\frac{2ig}{(2\pi)^{3}s}{\vec{\Delta}_{\perp}}^{2}
\int d^{2}\vec{r}_{\perp}V(\sqrt{\vec{r}_{\perp}^{2}+z'^{2}};s)]
e^{i{\vec{\Delta}_{\perp}}{\vec{r}_{\perp}}}\times$$
$$
\times\int_{-\infty}^{\infty} zdz
V(\sqrt{\vec{r}_{\perp}^{2}+z^{2}};s) \exp {\biggl[ \frac{2ig}{s}
\int_{z}^{\infty} dz'V(\sqrt{\vec{r}_{\perp}^{2}+z'^{2}};s)\biggl ]}.
\eqno(4.10)
$$
In Eq.(4.10) the first term describes the leading
eikonal behavior of the scattering amplitude, while the remaining
terms determine the corrections of relative magnitude $1/\sqrt{s}$. Due to the smoothness of the potential $V$ at high energy $s\rightarrow\infty$ the change of the particle momentum $\vec{\Delta}_\bot$, is relatively small. Therefore, the terms proportional to $\vec{\Delta}_\bot V$ and $\vec{\Delta}_\bot^2 V$ in Eq.(4.10) can be neglected, now we have \\
$$
{\left.{T_{Scalar}^{(0)}(s;t)} \right|_{\begin{array}{*{10}{c}}
{s\to \infty }\\
{t - fixed}
\end{array}}} = -\frac{is}{{2{{\left( {2\pi}\right)}^3}}}\int {{d^2}{r_ \bot }{e^{i{{\vec \Delta }_ \bot }{{\vec r}_ \bot }}}} \left\{ {\exp \left[ {\frac{{2ig}}{s}\int_{-\infty}^{\infty}{dzV\left( {\sqrt {{{\vec r}^2}_ \bot  + {z^2}} ;s} \right)} } \right] - 1} \right\}
\eqno(4.11)
$$
$$T_{Scalar}^{(1)}{\left. {(s;t)} \right|_{\begin{array}{*{10}{c}}
{s \to \infty }\\
{t - fixed}
\end{array}}} =  - \,\frac{{6{g^2}}}{{{{\left( {2\pi } \right)}^3}s\sqrt s }}\int {{d^2}{{\vec r}_ \bot }{e^{i{{\vec \Delta }_ \bot }{{\vec r}_ \bot }}}} \exp \left[ {\frac{{2ig}}{s}\int_{-\infty}^{\infty}{dz V\left( {\sqrt {\vec r_ \bot ^2 + {{z}^2}} ;s} \right)} } \right]\times
$$
$$\times\int_{-\infty }^\infty {dz}V^2\left( {\sqrt {\vec r_ \bot ^2 + {z^2}} ;s} \right)
\eqno(4.12)
$$

\indent As it is well known from the investigation of the scattering amplitude in the Feynman diagrammatic technique, the high energy asymptotic behavior can contain only logarithms and integral powers of $ s $. A similar behavior is observed here, since the integration of Eq.(4.10) leads to vanishing coefficients for half-integral powers $ s $. Nevertheless, allowance for the terms that contain the half-integral powers $ s $ is needed for the calculations of the next corrections in the scattering amplitude, and leads to the appearance of the so-called retardation effects, which are absent within the approximation used in the principal asymptotic terms.\\

\indent For the first term from Eq.(4.10) in the limit of high energies $s\rightarrow\infty$ and for $t$ - ﬁxed momentum transfers, with the assumption of smooth behavior the smooth behavior of the quasi-potential  as a function of the relative coordinate of two “nucleons”, in the framework of quantum field theory we find the leading eikonal of the high energy scattering amplitude\footnote{The amplitude $T$ is normalized in the c.m.s by the relation $\frac{d\sigma}{d\Omega}=\frac{|T(s,t)|^2}{64\pi^2s},\sigma_t=\frac{1}{2p\sqrt{s}}Im T(s,t=0)$}
$$
{\left.{T_{Scalar}^{\left( 0 \right)}(s;t)} \right|_{\begin{array}{*{10}{c}}
{s\to \infty }\\
{t - fixed}
\end{array}}}
= -\frac{is}{2(2\pi)^3}\int d^2 r_\bot e^{i\vec{\Delta}_\bot \vec{r}_\bot} \left\{\exp\left[i\chi_0(|\vec{r}^2_\bot|;s)\right]- 1\right\}
\eqno(4.13)
$$
$$
\chi_0\left(|\vec{r}_\bot|;s\right)=-\frac{g^2}{(2\pi)^2s}{K_0}\left(\mu|\vec{r}_\bot|\right)
$$
where $|\vec{r}_\bot|$ is a two-dimensional vector perpendicular to the nucleon – collision direction (the impact parameter), $K_0(\mu|\vec{r}_\bot|)$  is Mac Donald function of zeroth order, $\mu$ is a graviton mass which serves as an infrared cut-off and $\chi_0(|\vec{r}_\bot|;s)$ is the eikonal phase function.\\
\indent The similar leading eikonal Eq.(4.13) for the Lagrangian interaction $L_{int} = g\varphi^2(x)\phi(x)$ is also found by means of the functional integration[14], where eikonal phase function for exchange scalar virtual  meson corresponding to a Yukawa interaction potential between two “nucleons” $V(|\vec{r}_\bot|;s)= -\left(\frac{g^2}{4\pi s}\right)\left( \frac{e^{-\mu |\vec {r}_\bot|}} {\vec {r}_\bot}\right)$[14]. Use this exact form of the Yukawa as a quasi-potential and replace into Eqs. (4.11) and (4.12),  for  the leading eikonal amplitude and its first correction, we get (see appendix C)
$$
T_{Scalar}^{(0)}(s;t)
= \frac{{{g^2}}}{{2{{\left( {2\pi } \right)}^4}s}}.\left[ {\frac{1}{{{\mu ^2} - t}} - \frac{{{g^4}}}{{4{{\left( {2\pi } \right)}^2}{s^2}}}{F_1}(t) + \frac{{{g^7}}}{{12{{\left( {2\pi } \right)}^5}{s^4}}}{F_2}(t)} \right]
\eqno(4.14)
$$
$$
T_{Scalar}^{(1)}(s;t)
=\frac{{3i{g^6}}}{{4{{\left( {2\pi } \right)}^6}}}\frac{1}{{{s^3}\sqrt s }}\left[ {{F_1}(t) - \frac{{{g^3}}}{{{{(2\pi )}^3}{s^2}}}{F_2}(t)} \right]
\hspace{1cm}
\eqno(4.15)
$$

where
$$
F_1(t)= \frac{1}{t\sqrt{1-\frac{4\mu^2}{t}}}ln\left|
\frac{1-\sqrt{1-4\mu^2/t}}{1+\sqrt{1-4\mu^2/t}}\right|\eqno(4.16)
$$
and
$$
F_2(t)=\int_0^1
dy\frac{1}{(ty+\mu^2)(y-1)}ln\left|\frac{\mu^2}{y(ty+\mu^2-t)}\right|\eqno(4.17)
$$
The similar calculations can be applied  for other exchanges with
different spins. In the case of the vector model $ L_{int}=
-g\varphi^{\star}i\partial_\sigma\varphi
A^{\sigma}+g^{2}A_{\sigma\sigma}A^{\sigma}\varphi\varphi^{\star}\varphi$
the quasi-potential is independent of energy $V(|\vec{r}_\bot|)=-(g^2/4\pi)(e^{-\mu |\vec{r}_\bot|}/|\vec{r}_\bot|)$, we get
$$
T^{(0)}_{Vector}(s,t)=\frac{{{g^2}}}{{2{{\left( {2\pi } \right)}^4}}}.\left[ {\frac{1}{{{\mu ^2} - t}} - \frac{{{g^4}}}{{4{{\left( {2\pi } \right)}^2}s}}{F_1}(t) + \frac{{{g^7}}}{{12{{\left( {2\pi } \right)}^5}{s^2}}}{F_2}(t)} \right]\eqno(4.18)
$$
$$
T^{(1)}_{Vector}(s,t)=\frac{{3i{g^6}}}{{4{{\left( {2\pi } \right)}^6}}}\frac{1}{{s\sqrt s }}\left[ {{F_1}(t) - \frac{{{g^3}}}{{{{(2\pi )}^3}s}}{F_2}(t)} \right]
\hspace{1cm}
\eqno(4.19)
$$
In the case of tensor model\footnote{The model of interaction of a
scalar "nucleons" with a gravitational field in the linear approximation to $
h^{\mu\nu}(x)L(x)=L_{0,{\varphi}}(x)+L_{0,grav.}(x)
+L_{int}(x)$, where
$$ L_{0}(x)=\frac{1}{2} [\partial^{\mu} \varphi (x)\partial_{\mu} \varphi(x)
-m^{2} {\varphi}^{2}(x)],$$
$$ L_{int}(x)=-\frac{\kappa}{2} h^{\mu\nu}(x)T_{\mu\nu}(x),$$
$$ T_{\mu\nu}(x)=\partial_{\mu}\varphi(x) \partial_{\nu}\varphi(x)-
\frac{1}{2}\eta_{\mu\nu} [\partial^{\sigma} \varphi
(x)\partial_{\sigma} \varphi(x) -m^{2} {\varphi}^{2}(x) ],$$
$T_{\mu\nu}(x)$ is the energy momentum tensor of the scalar field. The
coupling constant $\kappa$ is related to Newton's constant of
gravitation $G$ by $\kappa^2=32\pi G=32\pi l_{PL}^2, l_{PL}=1,6.10^{-33}cm$ is the Planck length},
the quasi- potential increases with energy
$$V(|\vec{r}_\bot|;s)= (\kappa^2 s/4\pi)\left(e^{-\mu|\vec{r}_\bot|}/|\vec{r}_\bot|\right),$$
we have
$$
T^{(0)}_{Tensor}(s,t)=\frac{{{\kappa ^2}s}}{{{{\left( {2\pi} \right)}^4}}}.\left[ {\frac{1}{{{\mu ^2} - t}} - \frac{{{\kappa ^4}}}{{2{{\left( {2\pi } \right)}^2}}}{F_1}(t) + \frac{{{\kappa ^7}}}{{3{{\left( {2\pi } \right)}^5}}}{F_2}(t)} \right]\eqno(4.20)
$$
$$
T^{(1)}_{Tensor}(s,t)=\frac{{3i{\kappa ^6}\sqrt s }}{{{{\left( {2\pi } \right)}^6}}}\left[ {{F_1}(t) - \frac{{2{\kappa ^3}}}{{{{(2\pi )}^3}}}{F_2}(t)} \right]
\hspace{1cm}\eqno(4.21)
$$
Comparison of these above potentials has made it possible to draw the following conclusions: in the model with the scalar exchange, the total cross section $\sigma_t$ decreases as $\frac{1}{s}$, and only the Born term predominates in the entire eikonal equation; the vector model leads to a total cross section $\sigma_t$ approaching a constant value as $s\rightarrow\infty, \frac{t}{s} \rightarrow 0$. In both cases, the eikonal phases are purely real and consequently the inﬂuence of inelastic scattering is disregarded in this approximation, $\sigma^{in}=0$. In the case of graviton exchange the Froissart limit is violated. A similar result is also obtained in Ref.[11] with the eikonal series for reggeized graviton exchange.\\

\indent In the framework of the quasi-potential approach and the modified perturbation theory a systematic scheme of finding the leading eikonal scattering amplitudes and its corrections are developed and constructed in quantum field theory including the linearized gravity. The first correction to leading eikonal amplitude is found.
\section{The one loop approximation contribution to high energy  scattering}

\indent The low energy effective theory of quantized gravity is currently our most successful attempt at unifying general relativity and quantum mechanics [41-50]. Therefore, we attempt to extend the above approach to calculating the high energy scattering amplitude of two “nucleons” for the graviton exchange based on the Newtonian potential with low-energy leading one loop gravitational corrections anticipated by means of dimensional analysis. The non-relativistic Newtonian potential with all one loop gravitational corrections between two masses by means of Feynman diagrams has the following form in configuration space [42,55]
$$
{V_{Newton}}(r) =  - G\frac{{{m_1}{m_2}}}{r}\left[ {1 + 3\frac{{G\left( {{m_1} + {m_2}} \right)}}{{{c^2}r}} + \frac{{41}}{{10\pi }}\frac{{G\hbar }}{{{c^3}{r^2}}}\,} \right]
\eqno(5.1)
$$
which includes the lowest-order relativistic correction, and the lowest order quantum correction (also relativistic). The space parts of the non-analytical terms Fourier transform as
$$
\int {\frac{{{d^3}q}}{{{{(2\pi )}^3}}}} {e^{ - i\vec q\vec r}}\frac{1}{{{{\left| {\vec q} \right|}^2}}} = \frac{1}{{4\pi r}};
\int {\frac{{{d^3}q}}{{{{(2\pi )}^3}}}} {e^{ - i\vec q\vec r}}\frac{1}{{\left| {\vec q} \right|}} = \frac{1}{{2{\pi ^2}{r^2}}};
\int {\frac{{{d^3}q}}{{{{(2\pi )}^3}}}{e^{ - i\vec q\vec r}}\ln \left( {\frac{{{{\left| {\vec q} \right|}^2}}}{{{\mu ^2}}}} \right)=}
- \frac{1}{{2\pi {r^3}}}
\eqno(5.2)
$$
so clearly these terms will contribute to the corrections $V_{\text{Newton}}(r)$.\\
When substituting the Newtonian potential (5.1) into Eqs.(4.11) and (4.12), the vertex factor for graviton exchange between to “nucleons” should be added by $V_{\text{Newton}}$ with $sV_{\text{Newton}}$[14,50], graviton still has a mass $\mu$ and perform some calculations for the leading eikonal behavior and the first correction of the scattering amplitude we find the following expressions.
$$T_{graviton}^{(0)}(s,t) = \frac{{{\kappa ^2}s}}{{{{(4\pi )}^4}}}\left( {\frac{1}{{{\mu ^2} - t}} - \frac{{{\kappa ^3}}}{{2.{{(32\pi )}^2}}}{F_1}(t) + \frac{{2{\kappa ^6}}}{{3.{{(16\pi )}^5}}}{F_2}(t)} \right)
$$
$$
- \frac{{6({m_1} + {m_2})}}{{{{(32\pi )}^3}{c^2}}}\frac{{{\kappa ^5}s}}{{\sqrt {{\mu ^2} - t} }} + \frac{{41{\kappa ^5}s\hbar }}{{80.{{(4\pi )}^5}{c^3}}}{F_2}(t)
\eqno(5.3)
$$
and
$$
T_{graviton}^{(1)}(s,t) =\frac{{3i{\kappa ^6}\sqrt s }}{{{{(8\pi )}^6}}}\left[ {{F_1}(t) - \frac{{2{\kappa ^4}}}{{{{(8\pi )}^3}}}{F_2}(t)} \right] + \frac{{9({m_1} + {m_2})}}{{4{{(8\pi )}^3}{c^2}}}\frac{{{\kappa ^6}\sqrt s }}{{\sqrt {{\mu ^2} - t} }} + \frac{{123{\kappa ^6}\hbar \sqrt s }}{{10{{(4\pi )}^5}{c^3}}}{F_2}(t)
\eqno(5.4)
$$
\indent From the expressions (5.3) and (5.4) above, we see that the leading eikonal term scattering amplitude  and its first correction term have the same structure including three terms: i/ The first term is the scattering amplitude by exchanging gravitons, which in its non-relativistic limit will be the Newton potential; ii/The second term is the relativistic correction for the scattering amplitude (the term containing $(m_1+m_2)$). This term corresponds to the non-analytic contribution because of exchanging gravitons. The relativistic correction term is explained as the "zitterbewegung" fluctuation when the distance between two interacting particles is shifted by one Compton wavelength [41,42]; iii/The last term (the term proportional to $\hbar$) is quantum correction, obtained from the contribution of the one loop diagram in the high energy scattering process.\\

\indent The quantum correction term found in the linear gravitational field corresponds to the local interaction that is related to the analytical properties of the scattering amplitude. The Newtonian potential and its quantum corrections are related to the non-locality of the quasi-potential; non-analytic terms are also related to the non-locality of the Newtonian potential.\\

\indent The lowest order approximation of the modified perturbation theory is the leading eikonal scattering amplitude with phase function defined by the smooth local quasi-potential Yukawa in linearized quantum gravity. Using the quasi-potential Yukawa corresponding to virtual graviton between two “nucleons” we obtain the first correction to leading eikonal amplitude.\\

\indent  In the effective quantum gravity, the Newtonian quasi-potential, which contains the relativistic, and quantum corrections, we found the one-loop leading eikonal scattering amplitude and its first correction to leading amplitude. The difference here of the non-analytic and nonlocal contributions, connected with nonlocal quasi-potential were discussed and included in the calculation of the behavior scattering amplitude.

\section{Conclusion}

\indent In the framework of the modified perturbation theory and the quasi – potential equation, a systematic scheme of finding the leading eikonal scattering amplitudes and its corrections in the one-loop gravitational approximation in quantum gravity are developed and constructed. The first non-leading corrections to leading eikonal amplitude are found.

\indent In the linearized gravitational theory, the interaction between of scattered “nucleons” by exchanging graviton, which corresponds to the smooth quasi-potential of the Yukawa type. The leading eikonal scattering amplitude and its first correction also found.

\indent In the framework of effective field theory, we obtained the expression for the scattering amplitude in Newtonian potential taking into account the contribution of relativistic and quantum corrections from the one-loop diagram. The difference with the above case includes: i/ relativistic correction terms calculated from non-analytical contributions and explained as a result of "zitterbewegung" fluctuations when the distance between particles is shifted one Compton wavelength; ii/ quantum correction terms related to Planck's constant $\hbar$ was also found.\\

\indent The contributions to the high energy scattering amplitude are divided into analytic contributions related to the locality and non-analytic contributions associated with the non-locality. This division is associated with two ways of describing particles in quantum mechanics and relativistic quantum mechanics in that the particle has mass $m$. If the particle has mass, it is not possible to the localization of the particle in a volume with linear dimensions less than  the Compton wave-length of the corresponding particle. For ultra-relativistic particles such as  light quanta for which $(m=0,v=c)$- the concept of the  coordinates  of  the  particle  in  the  usual  sense  of  the  word  is  completely meaningless. [56].

\indent The new results in this paper are that we have incorporated relativistic effects and quantum effects in gravitational scattering to clarify part of the connection between the effective quantum field theory of general relativity and relativistic quantum mechanics.

\section{Acknowledgements}
We are grateful to Profs. B. M. Barbashov,V. A. Efremov, A. B. Arbuzov, V. R. Goloskokov, V. P. Gerdt,  A.J. Silenko  for valuable discussions. NSH is also indebted  to  Prof. G. Veneziano for suggesting this problem and his encouragement, and  Prof. H. Fried for useful remarks related to the modified perturbation theory. This work was supported in part by the joint Institute Nuclear Research (JINR, Dubna), the  Hanoi University of Science and funded by Vietnam National Foundation for Science and Technology Development (NAFOSTED) under grant number 103.01-2018.42\\

\noindent {\bf Appendix A: The smoothness of the local quasi-potential}

\indent Here we demonstrate that the eikonal representation is a consequence of the assumption of the non-singular character of hadron interactions at high energies. We shall use a quasi-potential equation for the wave function of two particles in the coordinate representation which has the form of a nonlocal differential equation
$$
\left({{E^2} - {m^2} + {{\vec \nabla }^2}} \right)\psi \left( {\vec r} \right) = \frac{1}{{\sqrt {{m^2} - {{\vec \nabla }^2}} }}V\left( {\vec r,s} \right)\psi \left( {\vec r} \right)
\eqno(A.1)
$$
where $\vec{r}$ is the vector of the relative coordinate of two “nucleons”. Under the condition of non-singular or smooth behavior quasi-potential $V(\vec{r},s)$, the non-localized differential equation  Eq. (A.1) takes an effectively local form in the high-energy limit. Actually, let us look for a solution of Eq.(A.1) of the form
$$\psi \left( {\vec r} \right) = {e^{ipz}}\varphi \left( {\vec r} \right)\eqno(A.2)$$
where $\varphi \left( {\vec r} \right)$ is expected to be a slowly varying functions compared to $exp(ipz)$ and $E = \sqrt{\vec {p}^2 + m^2}$.\\
It can easily be shown that on a space of slowly varying functions in the high-energy limit $p\rightarrow\infty$
$${e^{ - ipz}}\frac{1}{{\sqrt {{m^2} - {{\vec \nabla }^2}} }}{e^{ipz}} \to p - i{\nabla _z} + O\left( {\frac{1}{p}} \right)\eqno(A.3)$$
or
$$e^{-ipz}\left( {\sqrt {{m^2} - {{\vec \nabla }^2}} } \right){e^{ipz}} \to \frac{1}{p} + O\left( {\frac{1}{{{p^2}}}} \right)\eqno(A.4)$$
Thus the function $\varphi \left( {\vec r} \right)$ obeys the equation
$$-2ip\frac{{\partial \varphi \left( {\vec r} \right)}}{{\partial z}} = \frac{1}{p}V\left({s,r}\right)\varphi\left(\vec r\right)\eqno(A.5)$$
which coincides with the one which follows from the local Klein-Gordon equation with effective potential $\frac{1}{p}V(\vec{r},s)$. As a result we have the eikonal representation of elastic scattering amplitude with eikonal phase function $\chi_0(|\vec{r}_\bot|,s)$ (4.13). It is important note that the nonlocal potential differs from the local by terms of order $O\left(\frac{1}{p}\right)$ in the high energy limit $p\rightarrow\infty$. It should be noted, shifting from describing non-relativistic quantum theory to relativistic descriptions requires changing the concept of the coordinates of individual particles according to mass values. If the particle has mass, it is not possible to locate a particle in space with a length smaller than the Compton wavelength of the corresponding particle. For limited relativistic particles - light quanta - the concept of the particle's coordinates in the usual sense is completely absent[56]. \\

\noindent {\bf Appendix B: Relationship between the operator and Feynman path methods}\\

\indent What actual physical picture may correspond to our result given by Eq.(4.10)? To answer this question, we establish the relationship between the operator and Feynman path methods, which treats the quasi-potential equation Eq.(3.3) in the language of functional integrals. The solution of this equation can be written in the symbolic form
$$\exp(W) = \frac{1}{{1 - gK\left[ {{{\left( { - i\vec \nabla-\vec k} \right)}^2}} \right]V\left( {\vec r} \right)}} \times 1$$
$$=-i\int\limits_0^\infty  {d\tau } \,{e^{i\tau \left( {1 + i\varepsilon } \right)}}\exp \left\{ { - i\tau gK\left[ {{{\left( { - i\vec \nabla  - \vec k} \right)}^2}} \right]V\left( {\vec r} \right)} \right\} \times 1\eqno(B.1)$$
In accordance with the Feynman parametrization, we introduce an ordering index $\eta$ and write Eq. (B.1) in the form
$$\exp(W)=- i\int\limits_0^\infty  {d\tau } \,{e^{i\tau \left( {1 + i\varepsilon } \right)}}\exp \left\{ { - ig\int\limits_0^\infty  {d\eta K} \left[ {{{\left( { - i{{\vec \nabla }_{\eta  + \varepsilon }} - \vec k} \right)}^2}} \right]V\left( {{{\vec r}_\eta }} \right)} \right\} \times 1\eqno(B.2)$$
Using Feynman transformation
$$F\left[{P\left( \eta  \right)} \right] = \int {D\vec p} \int\limits_{x\left( 0 \right) = 0} {\frac{{D\vec x}}{{{{\left( {2\pi } \right)}^3}}}} \exp \left\{ {i\int\limits_0^\tau  {d\eta } \overrightarrow {\dot r}\left[ {\vec p\left( \eta  \right) - P\left( \eta  \right)} \right]} \right\}F\left[ {\vec p\left( \eta  \right)} \right]\eqno(B.3)$$
We write the solution of Eq. (2.6) in the form of the functional integral
$$\exp(W) =  - i\int\limits_o^\infty  {d\tau } {e^{i\tau \left( {1 + i\varepsilon } \right)}}\int {D\vec p} \int\limits_{x\left( 0 \right) = 0} {\frac{D\vec x}{(2\pi)^3}}\times$$
$$\times\exp \left\{ {i\int\limits_0^\tau  {d\eta } \overrightarrow {\dot x} \left( \eta  \right)\left[ {\vec p\left( \eta  \right) - P\left( \eta  \right)} \right]} \right\}G\left( {\vec x,\vec p;\tau } \right) \times 1\eqno(B.4)$$
In Eq.(B.4) we enter the function $G$
$$G\left( {\vec x,\vec p;\tau } \right) = \exp \left\{ { - i\int\limits_0^\tau  {d\eta } \overrightarrow {\dot x} \left( \eta  \right){\nabla _{\eta  + \varepsilon }}} \right\}\exp \left\{ { - ig\int\limits_0^\tau  {d\eta } K\left[ {{{\left( {\vec p\left( \eta  \right) - \vec k} \right)}^2}} \right]V\left( {{{\vec r}_\eta }} \right)} \right\}\eqno(B.5)$$
Which satisfies the equation
$$\frac{{dG}}{{d\tau }} = \left\{ { - igK\left[ {{{\left( {\vec p\left( \tau  \right) - \vec k} \right)}^2}} \right]\;V\left( {\vec r - \overrightarrow {\dot x} \left( {\tau  - \varepsilon } \right)} \right)\vec \nabla } \right\}G;{\rm{    }}G\left( {\tau  = 0} \right) = 1\eqno(B.6)$$
Finding from Eq.(B.6) the operator function $G$ and substituting it into Eq. (B.4) for $W$ we obtained the following final expression
$$\exp (W)=  - i\int\limits_o^\infty  {d\tau } {e^{i\tau \left( {1 + i\varepsilon } \right)}}\int {D\vec p} \int\limits_{x\left( 0 \right) = 0} {\frac{{D\vec x}}{{{{\left( {2\pi } \right)}^3}}}} \exp \left\{ {i\int\limits_0^\tau  {d\eta } \overrightarrow {\dot x} \left( \eta  \right)\vec p\left( \eta  \right)} \right\}\exp \left( {g\Pi } \right)\eqno(B.7)$$
where
$$\Pi=- i\int\limits_0^\tau  {d\eta } K\left[ {{{\left( {\vec p\left( \eta  \right) - \vec k} \right)}^2}} \right]V\left[ {\vec r - \int\limits_0^\tau  {d\xi \vec x\left( \xi  \right)\vartheta \left( {\xi  - \eta  + \varepsilon } \right)} } \right]\eqno(B.8)$$
$$\Pi^2=-i\int\limits_0^{{\tau _1}} {\int\limits_0^{{\tau _2}} {d{\eta _1}} } d{\eta _2}K\left[ {{{\left( {\vec p\left( {{\eta _1}} \right) - \vec k} \right)}^2}} \right]K\left[ {{{\left( {\vec p\left( {{\eta _2}} \right) - \vec k} \right)}^2}} \right]$$
$$\times V\left[ {{{\vec r}_1} - \int\limits_0^{{\tau _1}} {d\xi \vec x\left( \xi  \right)\vartheta \left( {\xi  - \eta  + \varepsilon } \right)} } \right] \times V\left[ {{{\vec r}_2} - \int\limits_0^{{\tau _2}} {d\xi \vec x\left( \xi  \right)\vartheta \left( {\xi  - \eta  + \varepsilon } \right)} } \right]\eqno(B.9)$$
Writing out the expansion
$$\overline{\exp \left( {g\Pi } \right)}= \exp \left( {g\bar \Pi } \right)\sum\limits_{n = 0}^\infty  {\frac{{{g^n}}}{{n!}}} \overline {{{\left( {\Pi  - \bar \Pi } \right)}^n}}\eqno(B.10)$$
in which the sign of averaging denoted integration with respect to $\tau,\vec{x}(\eta)$ and $\vec{p}(\eta)$ with the corresponding measure (see, for example Eq.(B.7) ), and performing the calculations, we find
$$W_1= \bar \Pi ,\quad {W_2} = \frac{1}{{2!}}\left( {\overline {{\Pi ^2}}  - {{\bar \Pi }^2}} \right),\quad {W_3} = \frac{1}{{3!}}\left[ {\overline {{\Pi ^3}}  - {{\bar \Pi }^3} - 3\bar \Pi \left( {\overline {{\Pi ^2}}  - {{\bar \Pi }^2}} \right)} \right]...\text{etc}\eqno(B.11)$$
the expressions (B.11) and (3.6), (3.7),(3.8) are identical.
$$W_1\left( {\vec r;\vec k;s} \right) = \bar \Pi  =  - i\int\limits_0^\infty  {d\eta } K\left[ {{{\left( {\vec p\left( \eta  \right) - \vec k} \right)}^2}} \right] \times \exp \left[ { - \int\limits_0^\tau  {d\xi \vec x\left( \xi  \right)\vartheta \left( {\xi  - \eta  + \varepsilon } \right)} } \right]V\left( {\vec r} \right)$$
$$=\int {d\vec q{e^{ - i\vec q\vec r}}} K\left[ {{{\left( {\vec q + \vec k} \right)}^2}} \right]V\left( {\vec q;s} \right),\hspace{3cm}\eqno(B.12)$$
$$\overline {{\Pi ^2}}  = K\left[ {{{\left( {{{\vec \nabla }_{{{\vec r}_1}}} + {{\vec \nabla }_{{{\vec r}_2}}} + \vec k} \right)}^2}} \right]K\left[ {{{\left( {{{\vec \nabla }_{{{\vec r}_1}}} + \vec k} \right)}^2}} \right] \times K\left[ {{{\left( {{{\vec \nabla }_{{{\vec r}_2}}} + \vec k} \right)}^2}} \right]V\left( {{{\vec r}_1};s} \right)V\left( {{{\vec r}_2};s} \right)$$
$$= \int {d{{\vec q}_1}} \int {d{{\vec q}_2}} {e^{ - i\left( {{{\vec q}_1} + {{\vec q}_2}} \right)\vec r}}K\left[ {{{\left( {\vec k + {{\vec q}_1} + {{\vec q}_2}} \right)}^2}} \right]\times\hspace{3cm}$$
$$\times\left\{ {K\left[ {{{\left( {\vec k + {{\vec q}_1}} \right)}^2};s} \right] + K\left[ {{{\left( {\vec k + {{\vec q}_2}} \right)}^2};s} \right]} \right\}V\left( {{{\vec r}_1};s} \right)V\left( {{{\vec r}_2};s} \right)\eqno(B.13)$$
$$W_2\left( {\vec r;\vec k;s} \right) =  - \frac{{W_1^2\left( {\vec r;\vec k;s} \right)}}{{2!}} + \frac{1}{2}\int {d{{\vec q}_1}} \int {d{{\vec q}_2}} V\left( {{{\vec q}_1};s} \right)V\left( {{{\vec q}_2};s} \right)\hspace{3cm}$$
$$\times \left\{ {K\left[ {{{\left( {{{\vec q}_1} + \vec k} \right)}^2};s} \right] + K\left[ {{{\left( {{{\vec q}_2} + \vec k} \right)}^2};s} \right]} \right\}{e^{ - i\left( {{{\vec q}_1} + {{\vec q}_2}} \right)\vec r}}\eqno(B.14)$$
$$W_3\left( {\vec r;\vec k;s} \right) =  - \frac{{W_1^3\left( {\vec r;\vec k;s} \right)}}{{3!}} + \int {d{{\vec q}_1}d{{\vec q}_2}d{{\vec q}_3}V\left( {{{\vec q}_1};s} \right)V\left( {{{\vec q}_2};s} \right)V\left( {{{\vec q}_3};s} \right)}\hspace{2cm}$$
$$\times K\left[ {{{\left( {{{\vec q}_1} + \vec k} \right)}^2};s} \right]K\left[ {{{\left( {{{\vec q}_2} + \vec k} \right)}^2};s} \right]K\left[ {{{\left( {{{\vec q}_3} + \vec k} \right)}^2};s} \right]{e^{ - i\left( {{{\vec q}_1} + {{\vec q}_2} + {{\vec q}_3}} \right)\vec r}}\eqno(B.15)$$
Restricting ourselves in the expansion (B.11) to the first term, we obtain the approximate expression (4.11) for the scattering amplitude, which corresponds to the allowance for the particle Feynman paths. These paths can be considered as a classical paths and coincide in the case of the scattering of high energy particles through small angles to straight-line paths trajectories.\\

\noindent {\bf Appendix C: Some Integrals Used in this paper}\\
Firstly, we consider the integral
$${I_1} = \int\limits_{ - \infty }^\infty  {dz} V\left( {\sqrt {r_ \bot ^2 + {z^2}} ,s} \right) =  - \frac{{{g^2}}}{{4\pi s}}\int\limits_{ - \infty }^\infty  {dz} \frac{{{e^{ - \mu \sqrt {r_ \bot ^2 + {z^2}} }}}}{{\sqrt {r_ \bot ^2 + {z^2}} }}
\eqno(C.1)
$$
Here $V\left( {r;s} \right) =  - \;\left( {{{{g^2}} \mathord{\left/{\vphantom {{{g^2}} {4\pi s}}} \right.
 \kern-\nulldelimiterspace} {4\pi s}}} \right)\left( {{{{e^{ - \mu r}}} \mathord{\left/{\vphantom {{{e^{ - \mu r}}} r}} \right.
 \kern-\nulldelimiterspace} r}} \right)$ is the Yukawa interaction potential between two “nucleons”.\\
Perform Fourier transform
$$
{I_1} =  - \frac{{{g^2}}}{{4\pi s}}\int\limits_{ - \infty }^\infty  {dz} \left( {\frac{1}{{{{(2\pi )}^3}}}\int {{d^3}k\frac{{{e^{ - \vec k\vec r}}}}{{{k^2} + {\mu ^2}}}} } \right) =  - \frac{{{g^2}}}{{2{{(2\pi )}^3}s}}\int {{d^2}{k_ \bot }\frac{{{e^{ - i{k_ \bot }{r_ \bot }}}}}{{k_ \bot ^2 + {\mu ^2}}}}  =  - \frac{{{g^2}}}{{2{{(2\pi )}^2}s}}{K_0}(\mu |{\vec r_ \bot }|)
\eqno(C.2)
$$
with ${K_0}(\mu |{\vec r_ \bot }| = \frac{1}{{2\pi }}\int {{d^2}{k_ \bot }\frac{{{e^{ - i{k_ \bot }{r_ \bot }}}}}{{k_ \bot ^2 + {\mu ^2}}}}$ is the Mac Donald of zeroth order.\\
The integral
$$
{I_2} = \int {{d^2}\left| {{{\vec r}_ \bot }} \right|} {e^{i{{\vec \Delta }_\bot }{{\vec r}_ \bot }}}{K_0}(\mu \left| {{{\vec r}_ \bot }} \right|) = 2\pi \int {d\left| {{{\vec r}_ \bot }} \right|\left| {{{\vec r}_ \bot }} \right|} {J_0}\left( {{{\vec \Delta }_ \bot }{{\vec r}_ \bot }} \right) = \frac{{2\pi }}{{{\mu ^2} + \Delta _ \bot ^2}} = \frac{{2\pi }}{{{\mu ^2} - t}}.
\eqno(C.3)
$$
The integral
$$
{I_3} = \int {{d^2}\left| {{{\vec r}_ \bot }} \right|} {e^{i{\Delta _ \bot }\left| {{{\vec r}_ \bot }} \right|}}K_0^2(\mu \left| {{{\vec r}_ \bot }} \right|) = \int {{d^2}q} \frac{1}{{{q^2} + {\mu ^2}}}\frac{1}{{{{(q + {\Delta _ \bot })}^2} + {\mu ^2}}}
\eqno(C.4)
$$
here, the result of the integral that obtained from calculating $I_2$ have been used.\\
Using method of Feynman parameter integral $\frac{1}{{ab}} = \int\limits_0^1 {dx} \frac{1}{{{{\left[ {ax + b(1 - x)} \right]}^2}}}$, we obtain
$$
{I_3} = ( - i\pi )\int\limits_0^1 {dx} \frac{1}{{\left[ {{\mu ^2} - tx(1 - x)} \right]}} = ( - i\pi ){F_1}(t)
\eqno(C.5)
$$
with the $F_1(t)$ - function is determined by the Eq.(4.16).\\
By the same method, the following integral is also calculated
$$
{I_4} = \int {{d^2}\left| {{{\vec r}_ \bot }} \right|} {e^{i{\Delta _ \bot }\left| {{{\vec r}_ \bot }} \right|}}K_0^3(\mu \left| {{{\vec r}_ \bot }} \right|)\hspace{6cm}
$$
$$
I_4 = \frac{1}{{{{(2\pi )}^2}}}\int {{d^2}{q_1}} \int {{d^2}{q_2}} \frac{1}{{q_1^2 + {\mu ^2}}}\frac{1}{{q_2^2 + {\mu ^2}}}\frac{1}{{{{({q_1} + {q_2} + {\Delta _ \bot })}^2} + {\mu ^2}}}=- \frac{1}{2}{F_2}(t)
\eqno(C.6)
$$
with the $F_2(t)$- function is determined by the Eq.(4.17).

\end{document}